\documentclass{tip}
\usepackage{mathtools}
\newcommand{\smallplus}{{\scriptscriptstyle +}}
\newcommand{\smallmin}{{\scriptscriptstyle -}}
\newcommand{\smallpm}{{\scriptscriptstyle \pm}}

\newcommand{\autora}{R.~Z\"ollner}
\newcommand{\autorb}{B.~K\"ampfer}
\title{\vspace{-2.5cm}\hrulefill\\\color{darkblue}\sffamily\LARGE\bfseries Two-Fluid Schwarzschild Solution\\}
\author{\sffamily\bfseries \autora${}^a$~\&~\autorb${}^{b,c}$}
\affil{\sffamily${}^a$Institut f\"ur Technische Logistik und Arbeitssysteme, TU~Dresden,\newline01062 Dresden, Germany}
\affil{\sffamily${}^b$Helmholtz-Zentrum  Dresden-Rossendorf, 01314 Dresden, Germany}
\affil{\sffamily${}^c$Institut f\"ur Theoretische Physik, TU~Dresden, 01062 Dresden, Germany}
\date{\sffamily\today\\\hrulefill}

\newcommand{\blau}[1]{\textcolor{black}{#1}}
\allowdisplaybreaks

\begin{document}
	\maketitle
	\thispagestyle{plain}
	\vspace{-2.0em}
	
	\parbox[t]{0.35\linewidth}{\textbf{Keywords}\linespread{1}\selectfont\raggedright
		\begin{itemize}
			\item compact stars \item core-corona decomposition \item interior two-fluid Schwarzschild solution \item dark matter
			\end{itemize}}
	\hfill
	\parbox[t]{0.55\linewidth}{\textbf{Abstract}\linespread{1}\selectfont\\We consider the interior two-fluid Schwarzschild solution. That is a model 
		of compact (neutron) stars with admixed dark matter. The two constant energy densities
		generalize the interior Schwarschild solution to the case of two fluids
		(representing standard model matter and dark matter, for instance), interacting mutually solely
		by the common gravitational field. In general, the two-fluid core is surrounded by a one-fluid corona (envelope).
		Only for special parameters (the model depends on energy densities $e_1,e_2$ and central pressures $p_{\mathrm{c}1},p_{\mathrm{c}2}$ of the fluids), both fluids occupy the same region.
		Despite the sum of two energy densities in that region, the compactness bound of \blau{4/9} is not exceeded. \blau{Therefore, a Buchdahl limit for two fluids holds.}
		A conceivable object is one with a stark leakage of one fluid (e.g.\ dark matter) beyond the two-fluid core.}
	
	\vspace{2.0em} 
	\noindent\rule{\textwidth}{0.4pt}
	\vspace{-3.0em}
	
\section{Introduction: The Quest for Dark Matter}\label{Sec1}
Dark matter (DM) is an element of contemporary astrophysics, introduced to solve a few puzzling facts
in cosmology and structure formation \cite{Cirelli:2024ssz,Garrett:2010hd,Balazs:2024uyj}.
Supposed DM is a manifestation of particles, it is expected to be captured and accumulated in compact objects,
such as white dwarfs and neutron stars. Given the fact that experiments aimed at observing DM particles
failed up to now, a wide-spread hypothetical spectrum of conceivable scenarios on the nature of them have been considered.
Among them are such ones of non-(self-)\ annihilating stable particles of various species 
with extremely small interaction cross section with standard model (SM) matter.
Accordingly, numerous models have been evaluated to investigate the impact on compact astrophysical objects. Neutron stars with SM and DM fluids are considered in~\cite{Iqbal:2026kjj,Dengler:2021qcq,Sotani:2025lzy,Shirke:2023ktu,Shakeri} particularly with fermionic and bosonic DM but strange DM and mirror DM as well. Models for boson stars can be found in~\cite{Pitz:2024xvh,Diedrichs:2023trk}. Cassing et al.~\cite{Cassing:2022tnn} consider two DM fluids. Gravitational wave events as constraints are used in~\cite{Wystub:2021qrn,Chen}, where~\cite{Yang:2024ycl} focuses on the available data to XTE~J1814-338. A two-shell (core-corona decomposition) is considered in~\cite{Zollner:2023myk,Zollner:2022dst,Cassing:2022tnn,Biesdorf:2024dor} for instance.
For surveys cf.\ \cite{Deliyergiyev:2019vti,Barbat:2024yvi,Grippa:2024ach}\blau{, and we refer the interested reader to the excellent textbook about compact star physics \cite{Buch}.}

Supposed DM has an exceedingly small direct interaction with SM matter,
one can argue that SM matter and DM
interact within neutron stars via their mutual gravitation fields. This idea leads to two-fluid models, where the
compact objects are treated  by the generalized Tolman-Oppenheimer-Volkoff (TOV) equations, when focusing on
spherical symmetry and static matter distribution.
Despite the numerous studies on DM in or surrounding neutron stars, a simple analytic base line model in the spirit
of the interior Schwarzschild solution is lacking~\cite{JBS}. 
Our note is aimed at filling this gap.
That is, we consider two fluids, each of constant energy density, and solve the two-fluid TOV equations. \blau{We find that a Buchdahl limit of 4/9 also applies for two fluids.}

Our paper is organized as follows. Section \ref{sec.II} presents the analytical solution of the two-fluid TOV equations.
The impact of DM on the pressure profile is considered in Section \ref{sec.III}. 
The special case of two fluids, uncovering the same spatial region, is dealt with in Section \ref{sec.IV},
to show that the Buchdahl limit \blau{for one fluid} is not exceeded.
The general case of \blau{different zeroes of the two pressure profiles leads to the situation that one of the fluids leaks out beyond the other one. Therefore, the outer shell of the compact object (so-called ``corona'' or ``envelope'') consists only of one fluid. These cases are} sketched in Section \ref{sec.V}. 
We summarize in Section \ref{sec.Summary}.

\section{Two-Fluid TOV Equations} \label{sec.II}

Let us consider two fluids, labeled by 1 and 2, which interact solely due to their mutual gravitational field.
The energy densities are supposed to be constant, $e_1$ and $e_2$, where one may imagine fluid 2 as DM.
The interior two-fluid Schwarzschild solution is governed by the two-fluid TOV equations in geometric units
(cf.\ \cite{Pitz:2024xvh} for instance\footnote{
	For the 1+1+2 covariant approach, see \cite{Naidu:2021nwh} partially based on~\cite{CarloniI,CarloniII}. Stability properties are dealt with
	in \cite{Caballero:2024qtv}.}):
\begin{alignat}{2}
	m_1' &= 4 \pi r^2 e_1, &&\quad m_2' = 4 \pi r^2 e_2, \label{eq.m12}\\
	p_1' &= -(e_1 + p_1) \Phi,  &&\quad p_2'= -(e_2 + p_2) \Phi, \label{eq.p12}\\
	\Phi &\mathrlap{\equiv \frac{m_1 + m_2  + 4 \pi r^3 p_1 + 4 \pi r^3 p_2}{r^2 \left( 1 - \frac{2 m_1}{r} - \frac{2 m_2}{r}\right) }
	= \frac{4 \pi}{3} r \frac{e^{\smallplus} + 3 p^{\smallplus}}{1 -  \frac{8 \pi}{3} e^{\smallplus} r^2},} &&\label{eq.Phi} 
\end{alignat}
where the prime denotes derivative w.r.t.\ the radial coordinate $r$,
and we denote $e^{\smallplus} \equiv e_1 + e_2$ and $p^{\smallplus} \equiv p_1 + p_2$.
Analogously, we define $e^{\smallmin} \equiv e_1 - e_2$ and $p^{\smallmin} \equiv p_1 - p_2$.
The quantities $m_{1 ,2}$ are the Misner-Sharp masses, and $p_{1, 2}$ stand for the pressures of
fluid 1 and 2.
The second equation in (\ref{eq.Phi}) uses $m_{1,2} = \frac{4 \pi}{3} e_{1,2} r^3$ (from (\ref{eq.m12}),
omitting a possible integration constant\footnote{Please note that a nonvanishing integration constant, i.e. ``central mass'', restricts the possible solutions such that the energy density and the central pressure cannot be chosen independently anymore.}). The interior two-fluid Schwarzschild solution reads
\begin{figure}
	\centering
	\includegraphics[width=0.45\columnwidth]{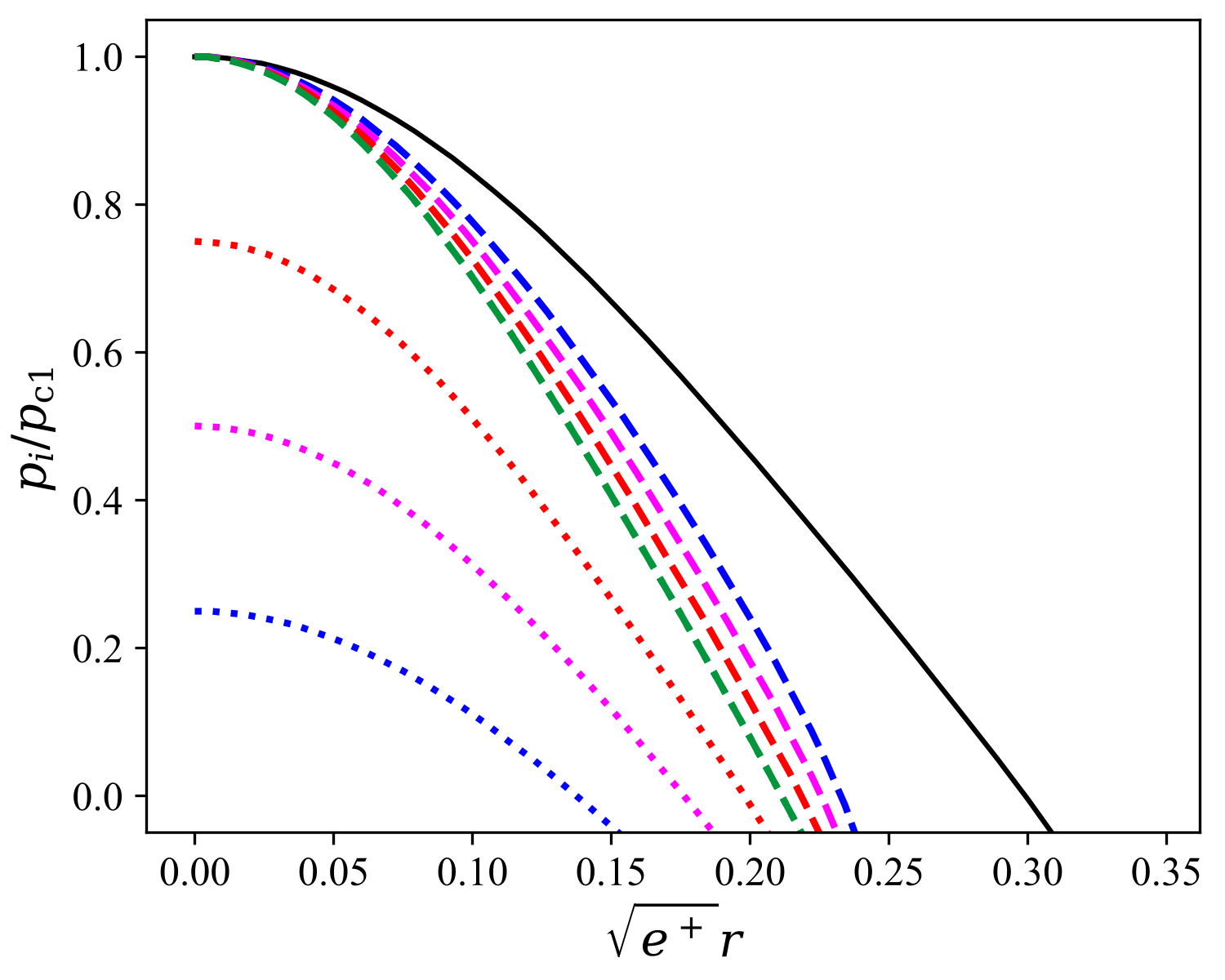} 
	\includegraphics[width=0.45\columnwidth]{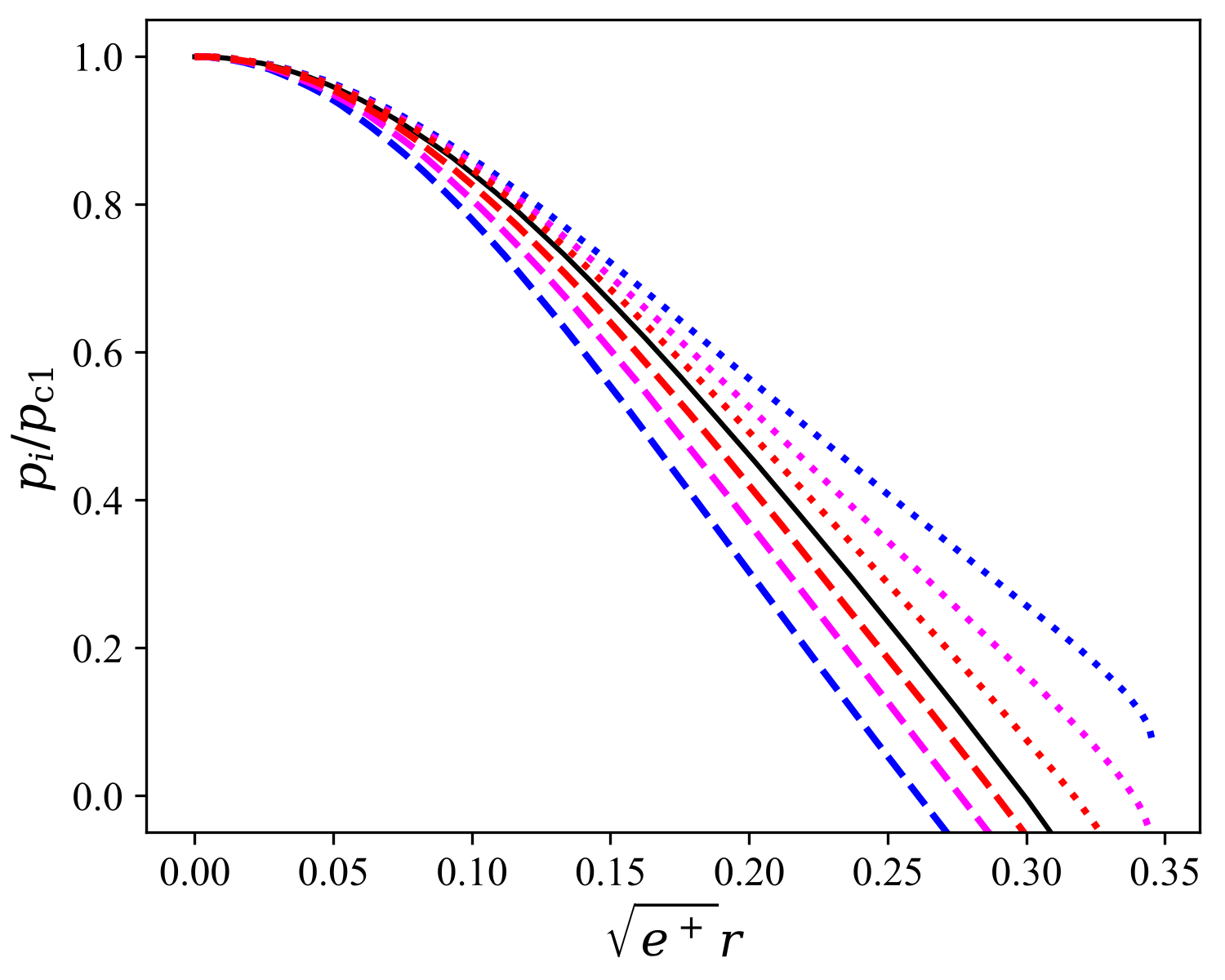}
	\caption{\blau{Pressure profiles $p_1$ (dashed, always with $e_1/p_{{\rm c}1}=1$) and $p_2$ (dotted) according to Equations~(\ref{eq.p1}, \ref{eq.p2}) as a function of the radial coordinate. The black solid curve corresponds to the (one-fluid) Schwarzschild solution. \textit{Left panel:} The energy density $e_2$ is kept constant ($e_2=p_{{\rm c}1}$), while $p_{{\rm c}2}/p_{{\rm c}1}$ runs from 0.25, 0.5, 0.75 to 1. \textit{Right panel:} Conversely, the central pressure $p_{{\rm c}2}$ is kept constant ($p_{{\rm c}2}=p_{{\rm c}1}$), while $e_2/p_{{\rm c}1}$ runs from 0.25, 0.5, 0.75 to 1 (analoguous coloring as left).}
		\label{fig0} 
	}
\end{figure}

\begin{align}
	p^{\smallplus}(r) &= - \frac{\mathcal{C} - W(r)}{3 \mathcal{C} - W(r)} e^{\smallplus} , \label{eq.p_plus}\\
	p^{\smallmin}(r) &= + \frac{3 \mathcal{C} - 1}{3 \mathcal{C} - W(r)} \left(e^{\smallmin} + p_{\mathrm{c}}^{\smallmin} \right) - e^{\smallmin} ,
	\label{eq.p_minus}
\end{align}
where dimensionless quantities are defined by
\begin{align}
	\mathcal{C} &\equiv \frac{e^{\smallplus} + p_{\mathrm{c}}^{\smallplus}}{e^{\smallplus} + 3 p_{\mathrm{c}}^{\smallplus}}, \label{eq.C}\\
	W(r) &\equiv \sqrt{1 - \frac{8 \pi}{3} e^{\smallplus} r^2}, \label{eq.W}
\end{align}
and the subscript ``c'' refers to the central pressure(s).
The pressures profiles follow as
\begin{align}
	2 p_1 (r) &= p^{\smallplus} + p^{\smallmin} = \frac{- (\mathcal{C} - W) e^{\smallplus} 
		+ (3 \mathcal{C} - 1) (e^{\smallmin} + p_{\mathrm{c}}^{\smallmin}) - (3 \mathcal{C} - W) e^{\smallmin}  }{3 \mathcal{C} - W} , \label{eq.p1}\\
	2 p_2 (r) &= p^{\smallplus} - p^{\smallmin} = \frac{ - (\mathcal{C} - W) e^{\smallplus} 
		- (3 \mathcal{C} - 1) (e^{\smallmin} + p_{\mathrm{c}}^{\smallmin}) + (3 \mathcal{C} - W) e^{\smallmin}  }{3 \mathcal{C} - W} . \label{eq.p2}
\end{align}
\blau{As a first glimpse, Figure~\ref{fig0} displays some examples according to Equations~(\ref{eq.p1}, \ref{eq.p2}).} These pressure profiles are linked by
\begin{align}
	p_1(r)= -e_1 + \frac{e_1 + p_{\mathrm{c}1}}{e_2 + p_{\mathrm{c}2}} ( e_2 + p_2(r)), \label{eq.p1(p2)}
\end{align}
that is, $p_1$ is a linear function of $p_2$.

If $e_1 p_{\mathrm{c} 2} = e_2 p_{\mathrm{c} 1}$, then the zero of $p_2(r)$ coincides with zero of $p_1(r)$ 
at the same value of the radial coordinate $r = R$.
Moreover, on this special hypersurface in parameter space (also called quadric cone)
$p_1 /p_2 = (e_1 + p_{\mathrm{c} 1})/(e_2 + p_{\mathrm{c} 2}) = p_{\mathrm{c} 1}/p_{\mathrm{c} 2}= const$; $const =1 $ is a further special case thereof.

If one is interested in the individual profiles $p_{1,2}(r)$ and the ordering of the pressure zeroes, then Equation~(\ref{eq.p1(p2)}) is useful too:
\begin{enumerate}[noitemsep,label=(\roman*)]
	\item $p_2(R) = 0 \Rightarrow$ $p_1 (R) > 0$ for $e_1 p_{\mathrm{c} 2} < e_2 p_{\mathrm{c} 1}$, 
	\item $p_2(R) = 0 \Rightarrow$ $p_1 (R) < 0$ for $e_1 p_{\mathrm{c} 2} > e_2 p_{\mathrm{c} 1}$. 
\end{enumerate}
Equation (\ref{eq.p1(p2)}) may be cast in the form
\begin{align}
	\frac{p_1(r)}{p_{\mathrm{c} 1}} = \frac{\delta}{(e_2 + p_{\mathrm{c} 2}) p_{\mathrm{c} 1}} 
	+ \frac{e_1  + p_{\mathrm{c} 1} }{e_2 + p_{\mathrm{c} 2}} \frac{p_{\mathrm{c} 2}}{p_{\mathrm{c} 1}} \frac{p_2(r)}{p_{\mathrm{c} 2}} ,
\end{align}
where $\delta \equiv  - e_1 p_{\mathrm{c}2} + e_2 p_{\mathrm{c}1}$.

As seen in Equation~(\ref{eq.W}), $W(r) \in [0,1]$ must be respected since $W(r=0) = 1$ and 
$W(r = r_{\mathrm{H}} \equiv \sqrt{3/(8 \pi e^{\smallplus}})) = 0$.
Accordingly, $p_1(W=1) = p_{c 1}$ and $p_2(W=1) = p_{\mathrm{c} 2}$.
Otherwise, 
\begin{align}
	p_1(W=0) &= \frac{e_2 (- e_1 + 2 p_{\mathrm{c} 1}) - e_1 (e_1 + p_{\mathrm{c} 1} + 3 p_{\mathrm{c} 2})}{3 (e_1 + e_2 + p_{\mathrm{c} 1} + p_{\mathrm{c} 2})}, 
	\label{eq.p1W0} \\
	p_2(W=0) &= \frac{- e_2 (e_1 +e_2 + 3 p_{\mathrm{c} 1} + p_{\mathrm{c} 2}) + 2 e_1 p_{\mathrm{c} 2}}{3 (e_1 + e_2 + p_{\mathrm{c} 1} + p_{\mathrm{c} 2})}.
	\label{eq.p2W0}
\end{align}
Observe the symmetry $1 \leftrightarrow 2$.
Equations (\ref{eq.p1W0}) and (\ref{eq.p2W0}) are noted in a form to highlight the impact of $e_2$ when keeping
the other three parameters: At small values of $e_2$, the leading orders are
$p_1 (W=0) \propto - e_1 (e_1 + p_{\mathrm{c} 1} + 2 p_{\mathrm{c} 2}) < 0$ and $p_2 (W=0)  \propto 2 e_1 p_{\mathrm{c} 2} > 0$
with common factors $(3 [e_1 + p_{\mathrm{c} 1} + p_{\mathrm{c} 2}])^{-1}$.
These symmetries are anchored in Equations~(\ref{eq.p1}, \ref{eq.p2}), where
$p_2$ emerges from $p_1$ by the replacements $e^{\smallmin} \to - e^{\smallmin}$ and $p_{\mathrm{c}}^{\smallmin} \to - p_{\mathrm{c}}^{\smallmin}$,
respectively $e_1 \to e_2$, $e_2 \to e_1$, $p_{\mathrm{c} 1} \to p_{\mathrm{c} 2}$, and $p_{\mathrm{c}2} \to p_{\mathrm{c} 1}$.

In contrast to the one-fluid Schwarzschild pressure profile, which extends from $p_{\mathrm{c}}$ to $- e_0/3$ (see below),
the profiles $p_{1 ,2}(r)$, respectively the profiles $p_{1, 2}(W)$, extend from $p_{\mathrm{c} 1, c 2}$ to $p_{1, 2}(W=0)$ given in
Equations~(\ref{eq.p1W0}, \ref{eq.p2W0}). Depending on the four parameters $\{e_1, e_2, p_{\mathrm{c} 1}, p_{\mathrm{c} 2} \}$
the latter values can be positive or negative. In the former case, this signals the possibility of an unstable configuration
since the pressure zero is not reached prior to the horizon $r_{\mathrm{H}}$.
For instance, let be the parameters  $\{e_1, p_{\mathrm{c} 1}, p_{\mathrm{c} 2} \}$ fixed and consider the dependence on $e_2$.
Then, $W_2 = 0$ (defined in Equation~(\ref{eq.W12}) below) is reached by $e_{2 \mathrm{crit}}$ given by
\begin{align}
	e_{2 \mathrm{crit}} = \frac12 \left(  -b + \sqrt{b^2-4 c}\right)  , \quad 
	b=e_1+2 p_{\mathrm{c}}^{\smallplus}+p_{\mathrm{c}}^{\smallmin}, \quad
	c=e_1 (-p_{\mathrm{c}}^{\smallplus}+p_{\mathrm{c}}^{\smallmin}) .
\end{align}


Given the above scalings, it is suitable to define a horizon by $2 m(r_{\mathrm{H}}) = r_{\mathrm{H}}$
introduced prior to Equation~(\ref{eq.p1W0}) 
and study the radial dependence of
$p_{1, 2} /p_{\mathrm{c} 1, 2}$ as a function of $r/r_{\mathrm{H}}$. 

The pressure zeroes are generically at $x_{1,2} = 1 - W_{1,2}^2$ with
\begin{align} 
	W_{1,2} =\frac{\mathcal{C} e^{\smallplus} \pm e^{\smallmin} \mp (3 \mathcal{C} - 1) p_{\mathrm{c}}^{\smallmin}}{e^{\smallplus} \pm e^{\smallmin}}
	\label{eq.W12}
\end{align}
or at $R_{1,2} = \sqrt{3 x_{1,2}/(8 \pi e^{\smallplus})}$. The scaled radial coordinate is here $x = \frac{8 \pi}{3} e^{\smallplus} r^2$.
One has separately to check whether $W_{1, 2} \in [0, 1]$.

The pressure profiles $p_1$ and $p_2$ cross each other at 
$W_\times = (e^{\smallmin} - (3 \mathcal{C} - 1) p_{\mathrm{c}}^{\smallmin})/e^{\smallmin}$,
where the corresponding radius $r_\times$ follows from Equation~(\ref{eq.W}).
One has separately to check whether $p_1 (r_\times) = p_2(r_\times) \ge 0$.
The special case $W_1=W_2 = W_\times=\mathcal{C}$ with  $p_1 (r_\times) = p_2(r_\times) = 0$ due to $\delta =0$
has been mentioned above.
Also here, one has separately to check whether $W_\times \in [0, 1]$.

The special setting $e_2 = 0$ and $p_{\mathrm{c} 2} = 0$, i.e.\ $e^{\smallplus} = e^{\smallmin} = e_1$ and
$p_{\mathrm{c}}^{\smallplus} = p_{\mathrm{c}}^{\smallmin} = p_{\mathrm{c} 1}$, leads to the Schwarzschild pressure profile.
Phrased differently, $e_2 = 0$ and $p_{\mathrm{c} 2} = 0$ in Equations~(\ref{eq.p1}, \ref{eq.p2})
brings back the Schwarzschild solution, 
\begin{equation}
p_1(r) = - e_1  \frac{\mathcal{C} - W(r)}{3 \mathcal{C} -W(r)} 
\end{equation}
and $p_2 (r) =0$.
Note that $W_{1,2}$ depend on the parameters $\{e^{\smallmin}/e^{\smallplus}, p_{\mathrm{c}}^{\smallmin}/e^{\smallplus}, p_{\mathrm{c}}^{\smallplus}/e^{\smallplus} \}$.
We select (somewhat arbitrarily) $e^{\smallplus}$ for scale setting in $W(r)$. 
The quantity $\mathcal{C}$ drops from $1$ at $p_{\mathrm{c}}^{\smallplus}/e^{\smallplus} = 0$ towards $1/3$ at $p_{\mathrm{c}}^{\smallplus}/e^{\smallplus}  \gg 1$.
In the latter limit, $W_{1,2} = \frac13 \frac{1 \pm 3 e^{\smallmin}/e^{\smallplus}}{1 \pm  e^{\smallmin}/e^{\smallplus}}$,
i.e.\ they run from the common point $1/3$ at $e^{\smallmin}/e^{\smallplus} = 0$ either to 0 at  $e^{\smallmin}/e^{\smallplus} = 1/3$ 
(lower branch) and $2/3$ at $e^{\smallmin}/e^{\smallplus} = 1$ (upper branch). Correspondingly, $x_{1,2}$
run from the common point $8/9$ at $e^{\smallmin}/e^{\smallplus} = 0$ to 0.75... (lower branch) and 1 (upper branch)
at $e^{\smallmin}/e^{\smallplus} = 1/3$. 

The pressures $p_{1, 2}(r)$ in Equations~(\ref{eq.p1}, \ref{eq.p2}), respectively $p^{\smallpm}(r)$ in
Equations~(\ref{eq.p_plus}, \ref{eq.p_minus}), may be positive and negative.
Before truncating the negative sections, we consider some examples. 

\section{Impact of Dark Matter on Pressure Profiles} \label{sec.III}

\begin{figure}[ht!]
	\centering
	\includegraphics[width=0.5\columnwidth]{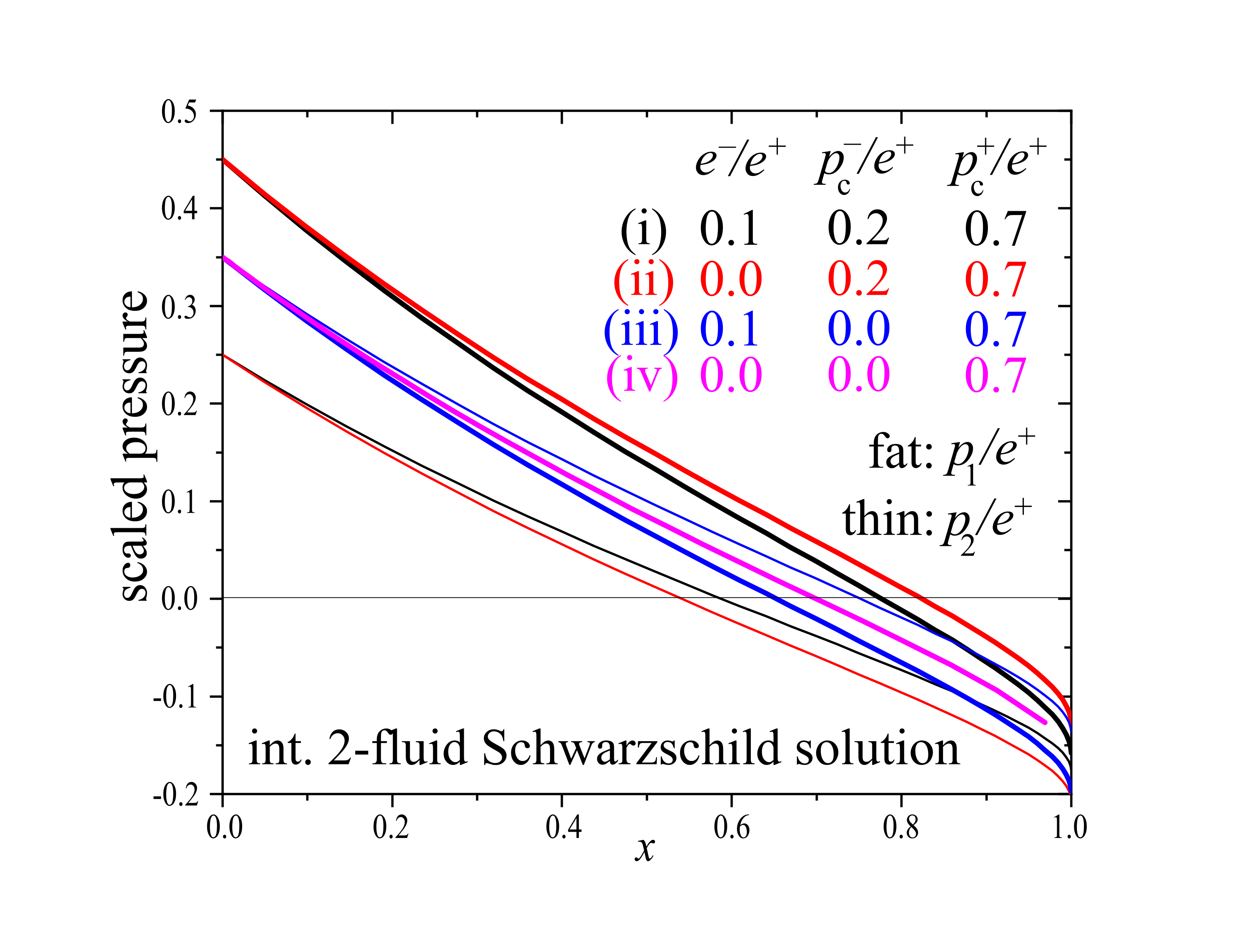}
	\caption{Four pairs of scaled pressure profiles $p_{1,2}/e^{\smallplus}$ of the interior two-fluid Schwarzschild solution
		Equations~(\ref{eq.p1}, \ref{eq.p2}) as a function of scaled radial coordinate
		$x = \frac{8 \pi}{3} e^{\smallplus} r^2$. For the parameters, see legend.
		Fat (thin) curves are for $p_1(r)/e^{\smallplus}$ ($p_2(r)/e^{\smallplus}$).
		In case of parameter choice (iv), $p_1 (r) = p_2 (r)$, i.e.\ fat and thin magenta curves are on top of another.
		\label{fig.2fluidSchwarz} 
	}
\end{figure}

Let us first get some impression of the impact of DM.
In doing so, the four-dimensional parameter space is constrained to a three-dimensional parameter space 
by using $e^{\smallplus}$ for a  scale setting. Accordingly,
$p_{1,2} /e^{\smallplus}$ depends on $ e^{\smallplus} r^2$ and parameters 
$p_{\mathrm{c}}^{\smallplus} /e^{\smallplus}, p_{\mathrm{c}}^{\smallmin}/e^{\smallplus}, e^{\smallmin}/e^{\smallplus}$.
Figure \ref{fig.2fluidSchwarz} exhibits the scaled  pressure profiles for four sets of parameters.
The case $p_{\mathrm{c}}^{\smallmin} = 0$ facilitates $p_1 (r=0) = p_2(r=0)$, see blue curves. However, due to $e^{\smallmin} \ne 0$,
the profiles split off. If additionally $e^{\smallmin} = 0$, then $p_1(r)$ and $p_2(r)$ are on top of another,
see magenta curve(s). 
Fat curves depict $p_1(r)/e^{\smallplus}$, while thin curves are for $p_2(r)/e^{\smallplus}$.
As stated above,  the sign of the combination $- e_1 p_{\mathrm{c} 2} + e_2 p_{\mathrm{c} 1} \equiv \delta$ determines the order
of the pressure zeroes. In the present examples, $p_2$ for cases (i, ii) has its zeroes first
($\delta = 0.065$, $\delta = 0.1$),
while in case (iii), $p_1$ drops first to zero ($\delta= - 0.035$).

\begin{figure}[ht!]
	\centering 
	\includegraphics[width=0.5\columnwidth]{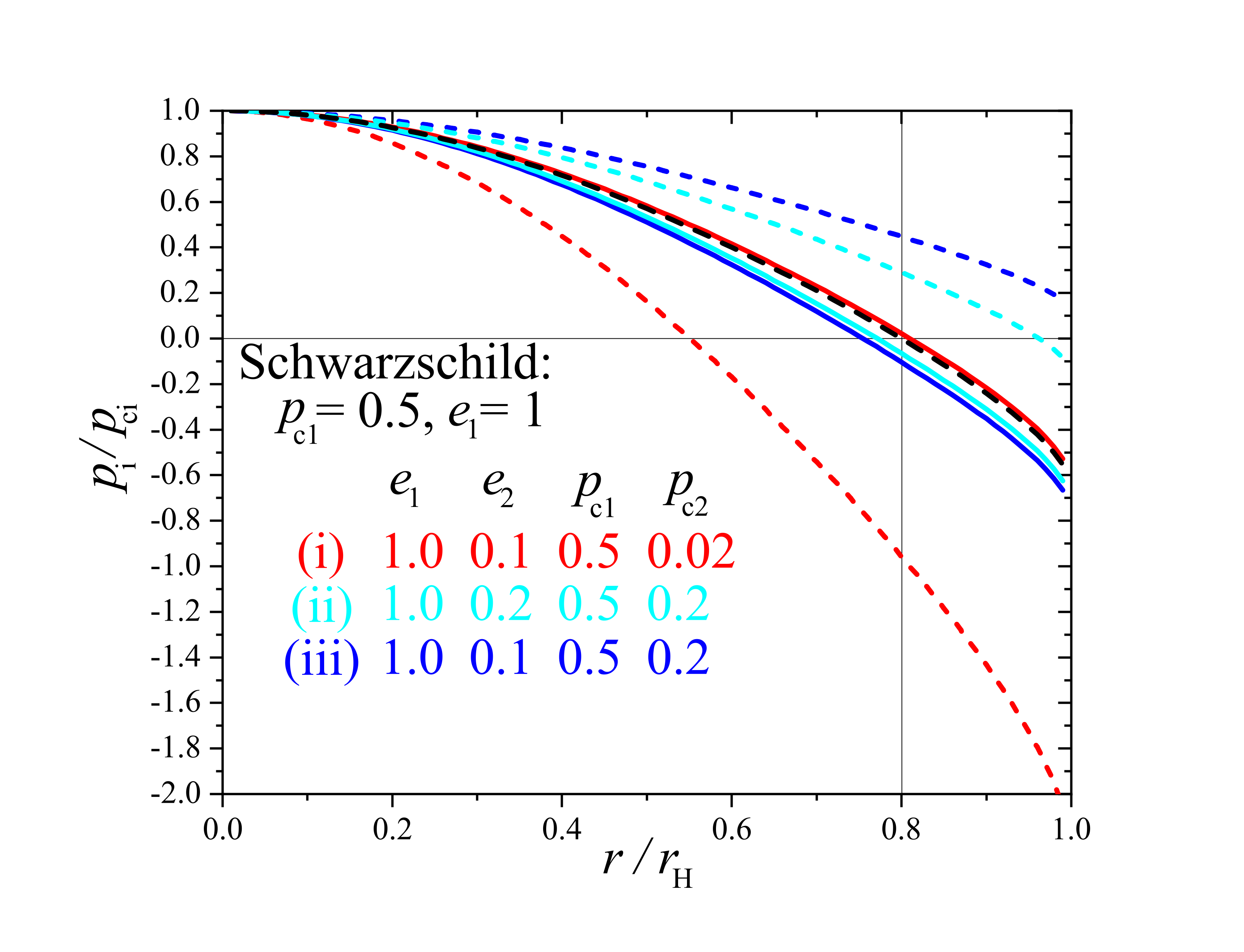}
	\caption{Scaled pressure profiles $p_{1,2}/p_{\mathrm{c} 1, \mathrm{c} 2}$ as a function of $r/r_{\mathrm{H}}$.
		For the selected parameters see legends and text.
		For comparison, the black \blau{dashed} curve (\blau{close to} the red curve) is for the Schwarzschild solution
		with $e_1 = 1$ and $p_{\mathrm{c} 1} = 0.5$.
		\label{fig.p(r_rH)} 
	}
\end{figure}

Another glimpse on the impact of DM on the two-fluid configuration is offered by a plot
of scaled pressures, $p_i/p_{\mathrm{c} i}~(i = 1, 2)$, as a function of $r/r_{\mathrm{H}}$. 
Note that $r_{\mathrm{H}}$ depends on both, $e_1$ and $e_2$.
Figure \ref{fig.p(r_rH)} exhibits such pressure profiles for
(i) $\{e_1, e_2, p_{\mathrm{c} 1}, p_{\mathrm{c} 2} \} = \{ 1, 0.1, 0.5, 0.02 \}$ (red curves),
(ii) $\{ 1, 0.2, 0.5, 0.2 \}$ (cyan curves) and (iii) $\{ 1, 0.1, 0.5, 0.2 \}$ (blue curves).\footnote{
	Remark on units: One should think on geometric units, 
	$r \propto \ell$, $m \propto \ell$, $e \propto \ell^{-2}$, $p \propto \ell^{-2}$,
	so that the combination $e r^2$ is dimensionless, as currently used.
	When attributing to $e$ or $p$ a dimensionless number, one keeps in mind that scaling 
	with the length or energy scale $\ell$. 
	The TOV equations (\ref{eq.m12}, \ref{eq.p12}, \ref{eq.Phi}) are independent of $\ell$.
}
$p_1/p_{\mathrm{c} 1}$ is exhibited by solid curves, while dashed curves are for $p_2/p_{\mathrm{c} 2}$ \blau{(or Schwarzschild)}.

To understand the parameter dependence let us cast Equations~(\ref{eq.p12}, \ref{eq.Phi}) in the form
\begin{align}
\frac{1}{p_{\mathrm{c} i}}	\frac{\mathrm{d} p_i}{\mathrm{d} \xi} = 
	- \frac12 \left( \left(\frac{p_{\mathrm{c} i}}{e_i} \right)^{-1} + \frac{p_i}{p_{\mathrm{c} i}} \right)
	\frac{\xi}{1 - \xi^2}  \left( 1 + 3 \frac{p_1 + p_2}{e_1 + e_2} \right) ,
\end{align}
where $\xi = r/r_{\mathrm{H}} = 0 \ldots 1$. For the considered parameters, the term $\left(\frac{p_{\mathrm{c}i}}{e_i} \right)^{-1}$
ranges from 5 to 1 to 0.5 for (i) to (ii) to (iii). That is, the gradient of $p_2$ in $\xi$ is largest for (i). Accordingly,
given the small value of $p_{\mathrm{c} 2}$, this scaled profile drops most rapidly, see red dashed curve.
In contrast, the blue dashed curve is flattest, owing the smallest value of  $\left(\frac{p_{\mathrm{c} i}}{e_i} \right)^{-1}$.
The term $p^{\smallplus}/e^{\smallplus}$ ranges from 0.47 to 0.58 to 0.63 for (i) to (ii) to (iii).
Accordingly, the scaled pressure gradients of $p_1$ change moderately, i.e.\ $p_1/p_{\mathrm{c} 1}$ drops somewhat
faster for (iii), see blue solid curve. Altogether, the deviations to the Schwarzschild case (depicted by the \blau{dashed} black curve)
are small. Measured by $\xi = r/r_{\mathrm{H}}$, the radii $R_1/r_{\mathrm{H}} = 0$ of $p_1 (R_1/r_{\mathrm{H}})$ shrink somewhat by the
impact of DM. For the selected parameters, the radii of the DM component change much more drastically.

The considerations above are not yet final, since they uncover positive and negative pressures.
The negative pressure branches must be truncated appropriately which requires more dedicated analyses. 

\section{Special Parameter Cases: Both Fluids Occupy the Same Region} \label{sec.IV}

Naively, one would expect that the compactness \blau{$\mathfrak{C} = M/R$} increases if two the energy densities
$e_1$ and $e_2$ occupy the same region enclosed in the radius $R_1 = R_2 = R$, since $M = \frac{4 \pi}{3} (e_1 + e_2) R^3$.
However, both fluids are coupled by the TOV equations which set the limit $2m_1(r) + 2 m_2(r) < r$ when
considering configurations with $p_{\mathrm{c} 1,2} < \infty$ and smoothly dropping pressure in going outward.

The condition for $R_1 = R_2$ was mentioned above,
$\delta = - e_1 p_{\mathrm{c} 2} + e_2 p_{\mathrm{c} 1} = 0$ or $e^{\smallmin} p_{\mathrm{c}}^{\smallplus} = e^{\smallplus} p_{\mathrm{c}}^{\smallmin}$.
Then $W_{1,2} = \mathcal{C}$ from Equation~(\ref{eq.W12}).
The compactness is accordingly \blau{
\begin{equation}
\mathfrak{C} = \frac{4 \pi}{3} (e_1 + e_2) R_1^2 = 2 X \frac{1 + 2X}{1 + 3X}
\end{equation}}
with $X \equiv \frac{p_{\mathrm{c} 1} + p_{\mathrm{c} 2}}{e_1 + e_2}$. Whatever the values of $e_{1,2} \ge 0$ and $p_{\mathrm{c} 1,2} \ge 0$
are (supposed $X \ge 0$), $\mathfrak{C}$ runs from zero (at small $X$) to \blau{4/9} (at large $X$), 
i.e.\ the Buchdahl limit as in the Schwarzschild case is recovered. \blau{Phrased differently, any two incompressible fluids on top of each other within a given radius can at most produce a compactness as for the one-fluid case. Therefore, the Buchdahl limit for one fluid holds for two fluids as well.}

\section{The Corona Surrounding a Two-Fluid Core} \label{sec.V}

The pressure zeroes follow from Equation~(\ref{eq.W12}).
If $W_1 = W_2$ due to $\delta = 0$, then no corona is needed. This is case (1) dealt with in the previous section.
If $W_1 \ne W_2$, one has to distinguish  two cases: \\
(2) $p_1(p_2=0) > 0$: $ \delta > 0$, and the corona consists of medium 1
with energy density $e_1$.\\
(3) $p_1(p_2=0) < 0$: $\delta < 0$, and the corona consists of medium 2
with energy density $e_2$.\\
Let us attribute the parameters $\{R_x,  M_x, p_x \}$ to the core radius, mass and surface pressure.
Then one integrates the one-fluid TOV equations, with energy density $e_x$, outward  up to vanishing pressure
with $M_x = \frac{4 \pi}{3} (e_1 + e_2) R_x^3$ and\\
(2) $R_x = R_2$,  $p_x =  \frac{  \delta}{e_2 + p_{\mathrm{c} 2}}$, $e_x = e_1$,\\
(3) $R_x = R_1$,  $p_x =  \frac{- \delta}{e_1 + p_{\mathrm{c} 1}}$, $e_x = e_2$. \\
In such a way the use of elliptic integrals mentioned below is avoided.

The reasoning behind that procedure is as follows.
For a viable star model, one has to continue beyond the first pressure zero 
(e.g.\ of fluid 2, as in the above cases (i) and (ii) in Figure~\ref{fig.2fluidSchwarz} or (i) in Figure~\ref{fig.p(r_rH)}) 
with respective zero energy density and pressure (e.g.\ $e_2 = 0, p_2 =0$). Beyond the second pressure zero, also the other energy density and pressure are set to be zero, i.e.\ the exterior Schwarzschild solution is to be applied.  
In other words, the core consisting of fluids 1 and 2, both with $p_{1,2} \ge 0$,
is surrounded by a one-fluid shell, the envelope or corona, which may be either fluid 1 or fluid 2,
depending on the chosen parameters, as exemplified in Figure~\ref{fig.2fluidSchwarz}.
The special case $p_1(r) = p_2(r)$ (this is case (iv) in Figure~\ref{fig.2fluidSchwarz}) does not require a corona. Instead,
outside of $p_1(R) = p_2(R) = 0$, the outer Schwarzschild solution with $M = m_1(R) + m_2(R)$ 
and $e_{1,2} = 0$, $p_{1,2} = 0$ applies at $r > R$ as emphasized in Section~\ref{sec.IV}.

Let us now assume, for definiteness, that $p_2 (R_2) = 0$ and $p_1 (R_1) =0$ with $R_2 < R_1$, as in cases (i) and (ii)
in  Figure~\ref{fig.2fluidSchwarz}. The opposite case (iii) corresponds to the \blau{dark} oyster \blau{introduced in} \cite{Biesdorf:2024dor}, \blau{where a relatively small core of SM matter is surrounded by a large halo of DM. This case is}
to be obtained by swapping the labels $1 \leftrightarrow 2$. We denote $m_1(R_2) = M_1$ and $p_1(R_2) = p_x$,
and analogously $m_2(R_2) = M_2$. The one-fluid corona at $r > R_2$ is determined by the TOV equations
\begin{align}
	m_1' &= 4 \pi r^2 e_1, \\
	p_1'  &= - \frac{(e_1 + p_1)(M_2 + \frac{4 \pi}{3} e_1 r^3 + 4 \pi r^3 p_1) }{r (r - 2 M_2 - \frac{8 \pi}{3} e_1 r^3)}. \label{eq.TOV2corona}
\end{align}
The terms with $M_2$ destroy the integration of Equation~(\ref{eq.TOV2corona}) by separation of variables.
Instead, it becomes a Bernoulli differential equation with the solution
\begin{equation}
	e_1+p_1(r) = \frac{(e_1+p_x)\omega}{\sqrt{1-\frac{2m}{r}}\left(1-( e_1+p_x) \omega\mathcal{I} \right) }
\end{equation}
with
\begin{align}
	\omega &\equiv\sqrt{1-\frac{8\pi}{3}e_1R_2^2},\\
	m(r) &\equiv M_2 + \frac{4\pi}{3}e_1r^3
\end{align}
and the integral
\begin{equation}
	\mathcal{I} \equiv 4 \pi \int\limits_{R_2}^{r} \frac{r\mathrm{d}r}{\left(1-\frac{2m}{r} \right)^{3/2} }.
\end{equation}
This integral can be expressed as a sophisticated combination of the elliptic integrals of the first and second kind. If one assumes $M_2$ to be sufficiently small, a perturbative expansion in $M_2$ leads to the approximation
\begin{equation}
	\mathcal{I} \approx \mathcal{I}_0 + M_2\mathcal{I}_1
\end{equation}
with 
\begin{align}
	\mathcal{I}_0 &= \frac{3}{2e_1} \left(1-\frac{8\pi}{3} e_1r^2\right)^{-\frac{1}{2}},\\
	\mathcal{I}_1 &= \frac{8 \pi r}{\left(1-\frac{8\pi}{3}e_1r^2 \right)^{\frac{3}{2}} } \left(\frac{3}{2} -\frac{8\pi}{3}e_1r^2\right).
\end{align}
The integration constants are fixed by the requirement $p_1(R_2) = p_x$. The total radius $R_{\mathrm{tot}}$ is to be found numerically by $p_1(R_{\mathrm{tot}}) =0$.
\\
\begin{figure}[ht!]
	\centering
	\includegraphics[width=0.45\columnwidth]{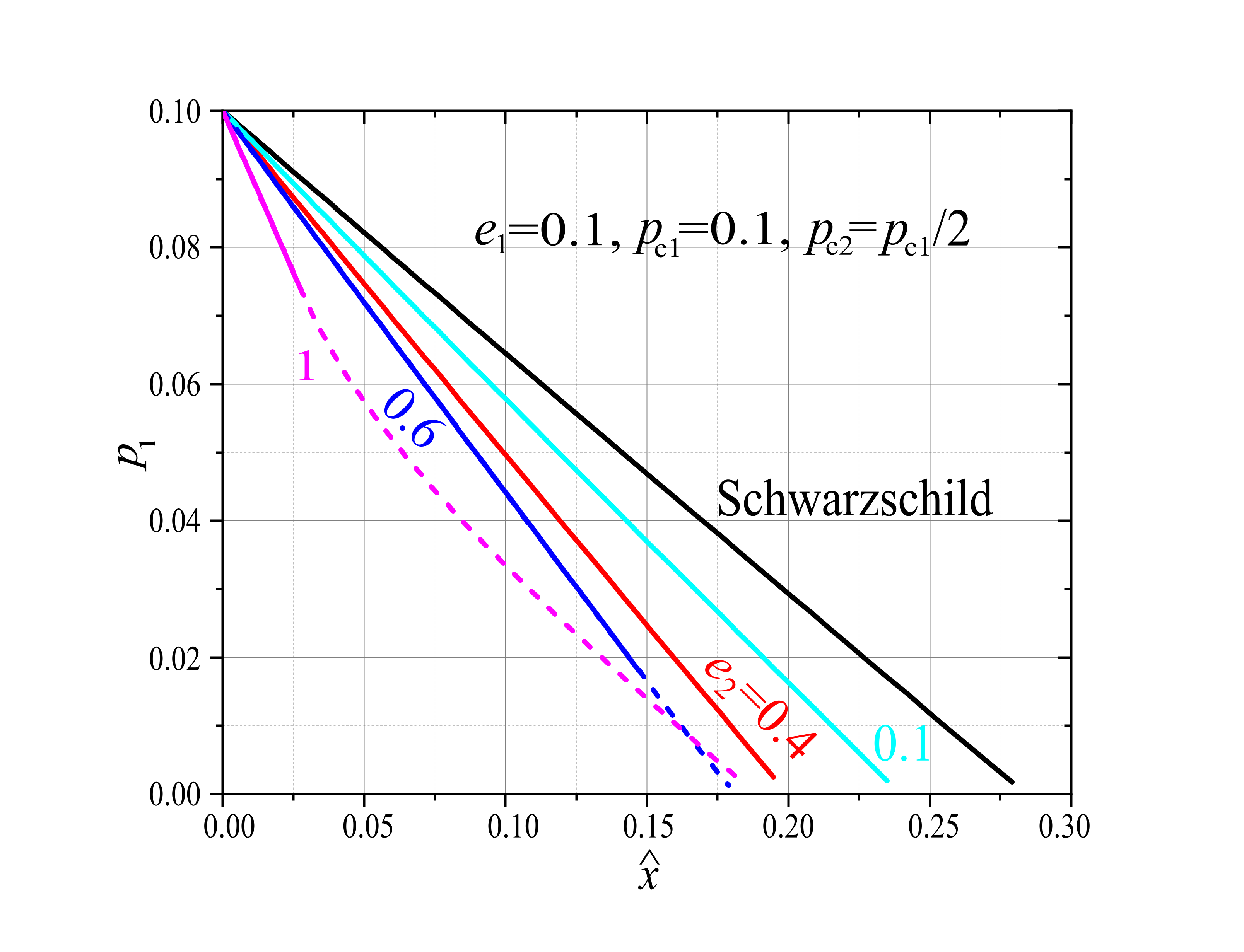} 
	\includegraphics[width=0.45\columnwidth]{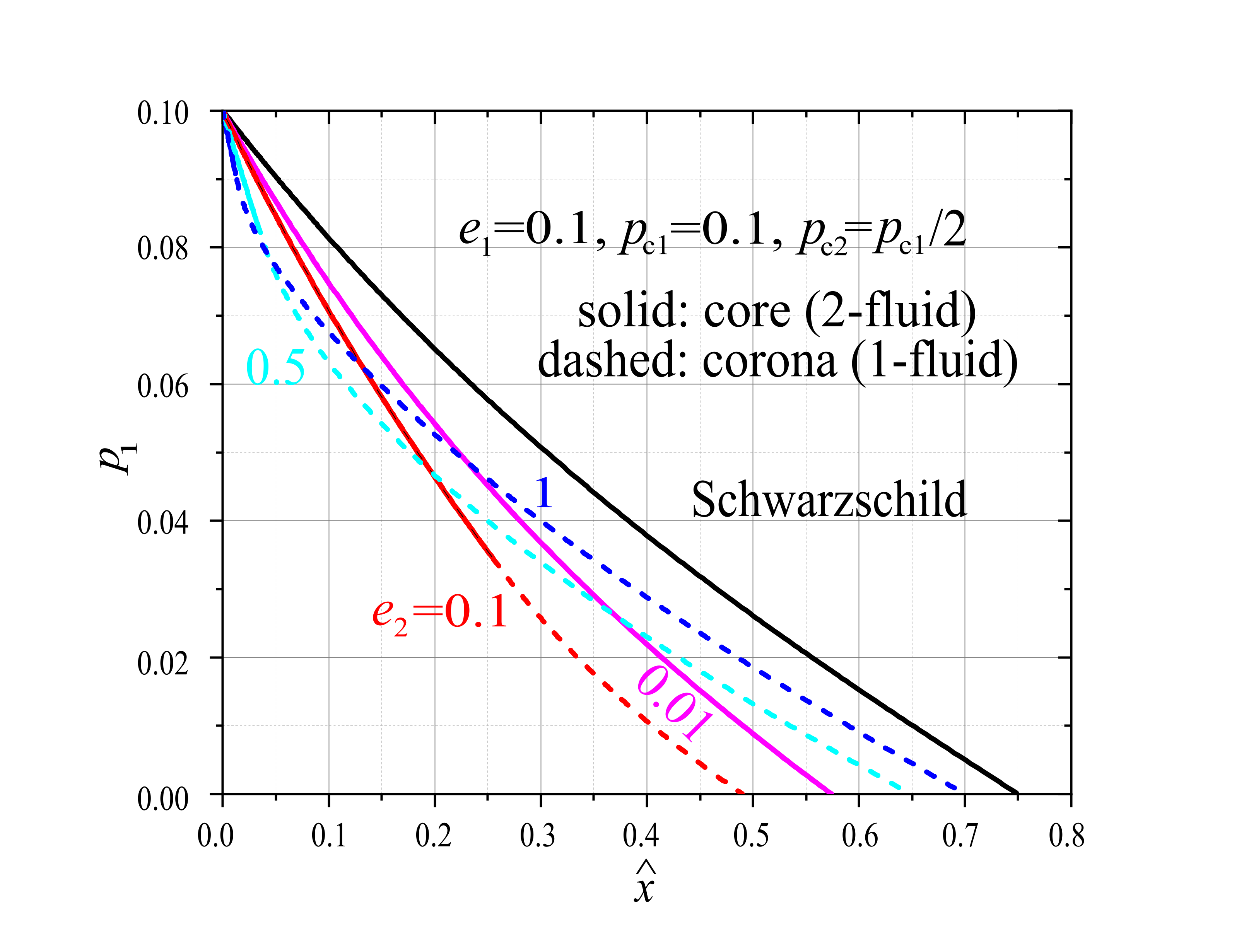}
	\caption{The pressure profile of $p_1$ as a function of $\hat x \equiv \frac{8 \pi}{3} e_1 r^2$
		with side conditions $p_1 \ge 0$ and $p_2 \ge 0$. The Schwarzschild solution is displayed by the
		solid black curves. Solid curves are for sections where $p_2 \ge 0$.
		The dashed curves continue $p_1$ with one-fluid Schwarzschild solution. 
		We have selected $p_{\mathrm{c} 1} = p_{\mathrm{c} 2}/2 = 0.1$.
		\textit{Left panel:} $e_1=1$  and varying values of $e_2 = 0.1, 0.4,0.6, 2$.
		\textit{Right panel:} $e_1=0. 1$, and varying values of $e_2 = 0.01, 0.1, 0.5, 1$.
		\label{fig.DMimpact} 
	}
\end{figure}
Figure \ref{fig.DMimpact} exhibits a few examples of the pressure profile $p_1(x)$ with 
dimensionless measure $\hat x \equiv \frac{8 \pi}{3} e_1 r^2$ of the radial coordinate $r$.
Various parameter combinations of $e_{1, 2}$ and $p_{\mathrm{c} 1, c 2}$ are chosen.
Those sections displayed by solid curves have $p_2 \ge 0$.
In cases where $p_2$ drops below zero at $p_1 > 0$, $e_2$ is put to zero in the TOV equations,
and the profile of $p_1$ is continued by the dashed sections.
For comparison, the Schwarzschild solution is exhibited.
The cyan and red curves in the left panel have $\delta < 0$, i.e. $p_2 > 0$ all the way of $p_1$
dropping from $p_{\mathrm{c} 1}$ to 0. The fluid 2 leaks out fluid 1 in such a case.
This is the \blau{dark} oyster case. The radius $R_2$ of the oyster is to be determined by $p_2(R_2) = 0$.
Changing $e_2$ from 0.4 to 0.6 changes the sign of $\delta$. The zero of the pressure $p_2$ is reached
at $p_1 > 0$ (displayed as solid section). At this zero one has to put $e_2 =0$ and continue the profile $p_1$
up to its zero (dashed section). In such case, the DM admixed core is inside the whole configuration.

The boundary case $\delta = 0$ is achieved by $e_2 = 0.5$, where $p_1 = 0$ and $p_2$ are reached simultaneously
(not displayed). This is why we selected the nearby values $e_2 = 0.4$ and 0.6. 

The right panel keeps $p_{\mathrm{c} 2} = p_{\mathrm{c} 1}/2$, but changes to $e_1 = 0.1$.
The value $e_2 = 0.01$ again corresponds to the \blau{dark} oyster, i.e.\ $p_1$ dropped to zero, while $p_2 > 0$ leaks out far away\footnote{Of course, the setting $p_{\mathrm{c} 2} \gg e_2$ looks somewhat weird, but 
	increasing $e_2$ turns $\delta$ to positive values letting shrink the DM radius coordinate.
	Nevertheless, the far out leaking DM halo for small $e_2$ is interesting in principle.}.
Increasing values of $e_2$ make the one-fluid corona larger, contrary to expectation
that a larger energy density makes the pressure gradient steeper. 

\begin{figure}[ht!]
	\centering
	\includegraphics[width=0.45\columnwidth]{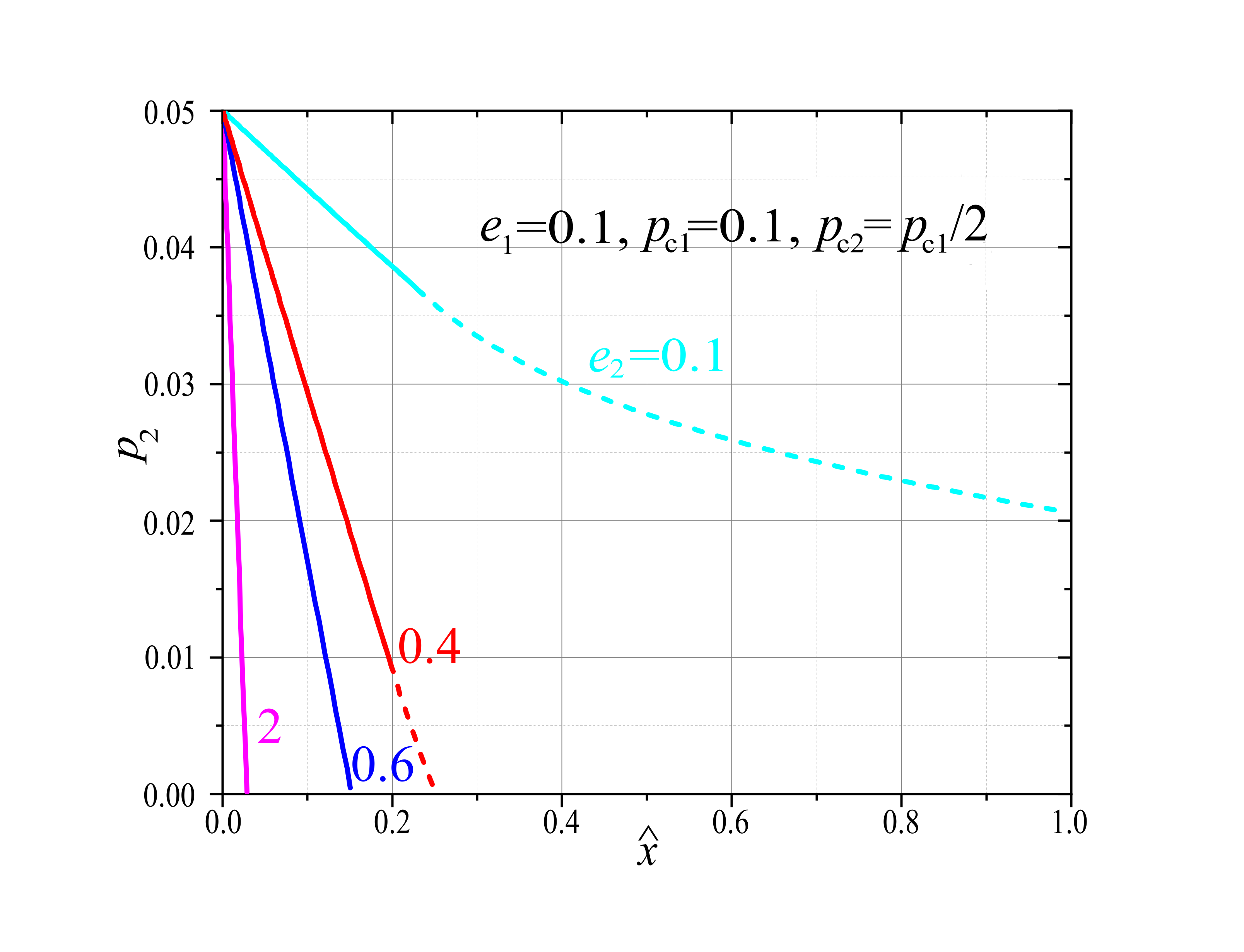} 
	\includegraphics[width=0.45\columnwidth]{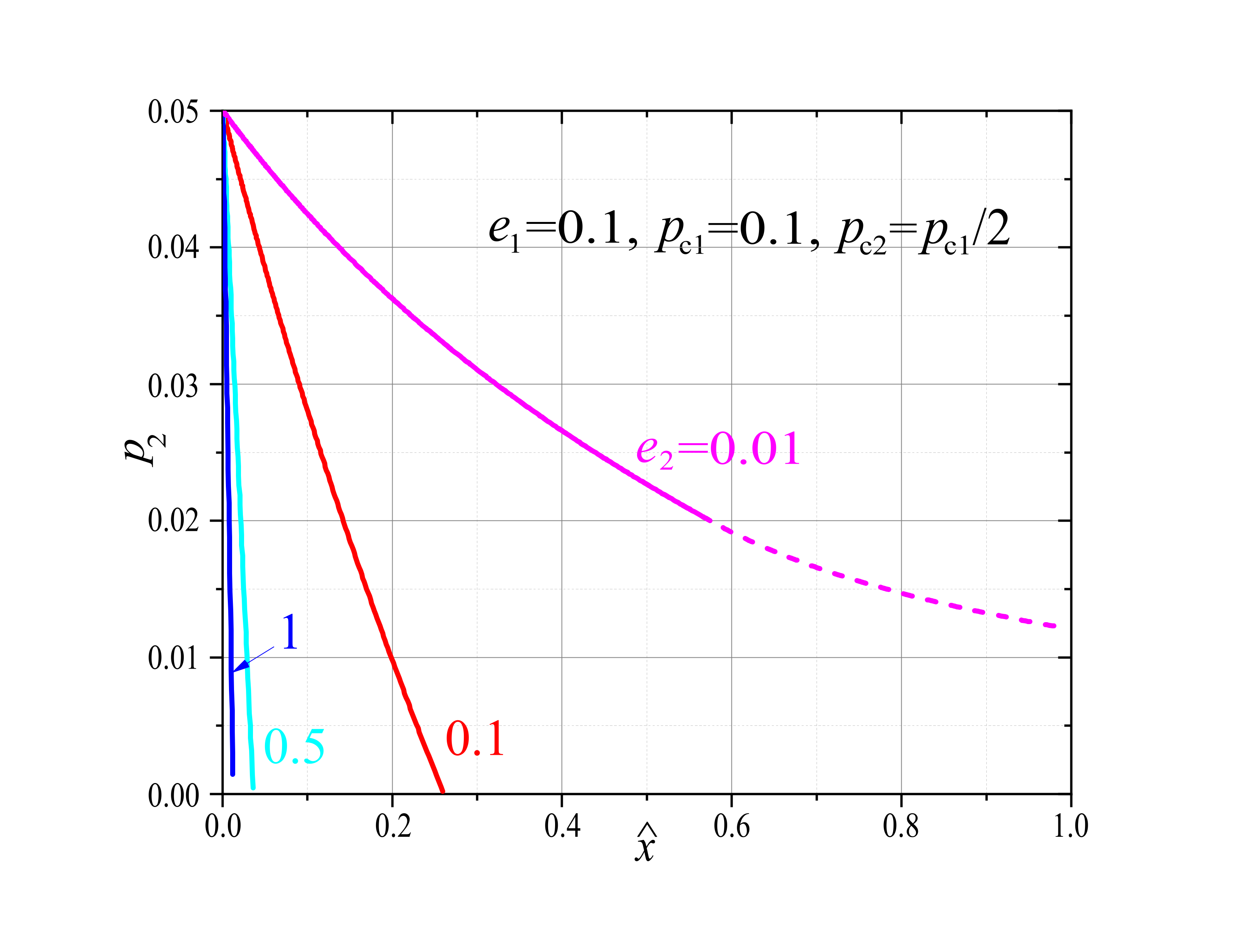}
	\caption{The pressure profile of $p_2$ as a function of $\hat x \equiv \frac{8 \pi}{3} e_1 r^2$
		analog to Figure~\ref{fig.DMimpact}. Note that, for the sake of comparison, the abscissa measure
		$\hat x$ is here used too.
		The side conditions $p_1 \ge 0$ and $p_2 \ge 0$ are imposed. 
		Solid curves are for sections where $p_1 \ge 0$.
		The dashed curves continue $p_2$ with one-fluid Schwarzschild solution,
		where $e_1 = 0$ and $p_1 = 0$.
		\label{fig.DMimpact_p2} 
	}
\end{figure}

The masses follow simply from $M_1 = \frac{4 \pi}{3} e_1 R_1^3$ and $M_2 = \frac{4 \pi}{3} e_2 R_2^3$.
The values of $R_1$ can be read off the figure, while $R_2$ needs the information on $p_2(x)$
given in Figure~\ref{fig.DMimpact_p2}. For the sake of comparison with  Figure~\ref{fig.DMimpact},
we employ here also the radial coordinate measure $\hat x = \frac{8 \pi}{3} e_1 r^2$.
In those cases where $\delta > 0$ (left panel with $e_2 = 0.6$ and 2, and right panel with $e_2 = 0.1, 0.5$ and 1),
the pressure $p_2$ drops to zero within the region where $p_1 > 0$, i.e.\ the corona consists of fluid 1.
Conversely, for those cases where $\delta < 0$ (left panel with $e_2 = 0.4$ and 0.1, and right panel with
$e_2 = 0.01$), the pressure $p_2$ drops to zero outside the fluid 1. That is, the corona consists of fluid 2,
thus forming the \blau{dark} oyster.
Particularly striking is the wide leakage of the corona like an extended halo for small values of $e_2$.
If one uses as measures of the radial coordinate $\bar x \equiv \frac{8 \pi}{3} e_2 r^2$, then the zero of $p_2$ is
seen at $\bar x < 1$. 
For parameter choices constrained to $p_{\mathrm{c} 2} < e_2$, the corona is not so extremely extended.
\section{Summary} \label{sec.Summary}

The famous interior Schwarzschild solution of Einstein's equations 
for a static fluid sphere with energy density $e_0$ and central pressure $p_{\mathrm{c}}$
(cf.\ \cite{Simmonds:2026jln} for a recent appraisal of the many facets)
delivers the pressure profile
\begin{equation}
	p(r) = - e_0 (\mathcal{C}_{0 c} - W_0(r))/(3 \mathcal{C}_{0 c} - W_0(r))
\end{equation}
with 
\begin{align}
	\mathcal{C}_{0 c} &\equiv (e_0 + p_{\mathrm{c}})/(e_0 + 3 p_{\mathrm{c}})\\
	W_0(r) &\equiv (1 - \frac{8 \pi}{3} e_0 r^2)^{1/2}
\end{align}
Considering the energy density $e_0 \ge 0$, the pressure runs from $p_{\mathrm{c}} = p(r=0) > 0$ to negative values
with increasing values of the circumferential radial coordinate $r$.
Defining the radius $R_0$ by the pressure zero, $p(R_0) = 0$, the Misner-Sharp mass becomes
$M = \frac{4 \pi}{3} e_0 R_0^3$ and serves as parameter of the exterior Schwarzschild solution.    
The present paper extends this one-fluid Schwarschild solution to the two-fluid Schwarzschild solution.
That is, two fluids 1 and 2 (with constant energy densities $e_{1, 2}$ and central pressures $p_{\mathrm{c} 1, c 2}$)
interact solely via their common gravitational field.
We present the profiles $p_{1, 2}(r)$ and their parametric dependencies.
In general, we face a four-dimensional parameter space.

Our focus is on the two-fluid core. 
In special cases, the pressure profiles of both components agree and the radius is determined by their agreeing 
pressure zeroes. This is the special case of a  quadric cone in the parameter space,
where both fluids occupy the same region, surrounded by vacuum.  
In general, however, the pressure zeroes of both fluids differ. Then, one defines the core radius by
that pressure zero with smaller radius, i.e.\ the first zero.
One has to continue the solution outside the core by a one-fluid zone (termed corona)
up to the second pressure zero. Depending on the parameters, the corona may consist of fluid 1 or fluid 2.
In the outer space region the famous exterior vacuum Schwarzschild solution applies.

It is surprising to see many studies of the impact of dark matter on properties of compact (neutron) stars.
What was missed in the literature is a comprehensive analytic reference dealing with the two-fluid Schwarzschild solution.
We fill this gap and emphasize the possible leakage of one fluid far beyond the two-fluid core.
Low energy density and high central pressure of dark matter can facilitate such an object (\blau{dark} oyster).  

Subsequent investigations should envisage the tidal deformability, stability properties and the formal aspects of
the sign of the energie densities $e_{1,2}$ and the central pressures $p_{\mathrm{c}1,\mathrm{c}2}$. Relations to metric quantities need to be considered.
Interesting are additionally the impact of energy conditions and the occurrence conditions of photon spheres. 
Also multi-layered structures should be considered to bridge over to realistic equations of state.


\subsection*{Acknowledgement}
	The authors gratefully acknowledge communications with J.~Schaffner-Bielich.

\begin{spacing}{1.0}

\end{spacing}
\end{document}